\documentclass[aps,prl,twocolumn,groupedaddress,showpacs]{revtex4-1}

\usepackage{graphicx}
\usepackage{bm}

\begin{document}


\title{Effects of three-nucleon spin-orbit interaction
on isotope shifts of Pb nuclei}


\author{H. Nakada$^1$}
\email[E-mail:\,\,]{nakada@faculty.chiba-u.jp}
\author{T. Inakura$^{1,2}$}
\affiliation{$^1$ Department of Physics, Graduate School of Science,
 Chiba University,\\
Yayoi-cho 1-33, Inage, Chiba 263-8522, Japan\\
$^2$ Yukawa Institute of Theoretical Physics, Kyoto University,\\
Kitashirakawa Oiwake-cho, Sakyo, Kyoto 606-8502, Japan}


\date{\today}

\begin{abstract}
We investigate effects of the $3N$ interaction,
which effectively adds a density-dependent term to the LS channel,
on the isotopes shifts of the Pb nuclei.
With the strength so as to keep the $\ell s$ splitting
of the single-nucleon orbits,
the density-dependence in the LS channel
tends to shrink the wave functions of the $j=\ell+1/2$ orbits
while makes the $j=\ell-1/2$ functions distribute more broadly.
Thereby the kink in the isotope shifts of the Pb nuclei at $N=126$
becomes stronger,
owing to the attraction from neutrons occupying $0i_{11/2}$ in $N>126$.
The density-dependence in the LS channel enables us
to reproduce the data of the isotope shifts
by the Hartree-Fock-Bogolyubov calculations
in a long chain of neutron numbers,
even without degeneracy between the $n1g_{9/2}$ and $n0i_{11/2}$ levels.
We exemplify it by the semi-realistic M3Y-P6 interaction.
\end{abstract}


\maketitle



\textit{Introduction.}
As finite quantum many-body systems dominated by the strong interaction,
atomic nuclei supply challenging problems for us.
One of the important questions is whether we need many-nucleon interaction
and, if so, how it affects the nuclear structure and reactions.
In particular, the three-nucleon ($3N$) interaction is under current interest.
The $3N$ interaction could become the more significant at the higher density,
because its energy contribution depends on the nucleon density more strongly
than that of the two-nucleon ($2N$) interaction.
This indicates astrophysical importance of the $3N$ interaction,
since energy of the nuclear matter at high density
is quite relevant to the supernovae and the neutron stars.
More and more evidence has been accumulated
from the detailed analysis of the few-nucleon systems~\cite{ref:3NF}
that the $3N$ interaction is not negligible.
It has also been argued that
the saturation properties extracted from the mass systematics
are difficult to be accounted for only by the $2N$ interaction,
and can be improved by the $3N$ interaction~\cite{ref:AK97}.
In connecting the results in the few-nucleon systems
to the saturation properties of the infinite nuclear matter
and the astrophysical problems,
it is desirable to reveal effects of the $3N$ interaction
in medium- to heavy-mass nuclei.
There have been arguments on structure of medium-mass nuclei,
\textit{e.g.} on the binding energies
in the O and Ca nuclei~\cite{ref:3NF-O&Ca}.
However, since it is not an easy task
to determine the $3N$ interaction with good precision,
its effects on nuclear structure are not always transparent
if arguments are based only on quantitative assessment of energies.
Effects of the $3N$ interaction on the nuclear structure
should be investigated further;
particularly observable effects on wave functions
of medium- to heavy-mass nuclei.

The spin-orbit ($\ell s$) splitting
in the nucleonic single-particle (s.p.) orbitals
is essential to the nuclear shell structure.
Whereas the size of the $\ell s$ splitting has been known from the data,
it has been questioned whether the $2N$ interaction
is able to produce sufficient $\ell s$ splitting~\cite{ref:AB81}.
The LS channels of the nucleonic interaction,
which contain $\mathbf{L}_{ij}\cdot(\mathbf{s}_i+\mathbf{s}_j)$
as will be given in Eq.~(\ref{eq:LS}),
determine overall size of the $\ell s$ splitting.
Recently Kohno has found~\cite{ref:Koh12,ref:Koh13}
that the $3N$ interaction from the chiral effective-field theory
($\chi$EFT)~\cite{ref:EGM05}
gives significant density-dependence in an LS channel
when it is converted to an effective $2N$ interaction,
and this sizably contributes to the $\ell s$ splitting.
To confirm whether this picture is correct,
it is intriguing to search effects of the density-dependence
on observables other than the $\ell s$ splitting.

The isotope shifts of the Pb nuclei,
in which a conspicuous kink has been observed at $N=126$
when plotted as a function of the neutron number $N$,
are significantly influenced by attraction from neutrons
occupying the $0i_{11/2}$ orbital~\cite{ref:RF95}.
Since the density-dependence in the LS channel
gives a broader $n0i_{11/2}$ function as illustrated below,
it may affect the isotope shifts.
In this paper we shall examine effects of the density-dependent LS channel
connected to the $3N$ interaction
on the isotope shifts of the Pb nuclei.

We mention that a density-dependent LS interaction
was considered in Ref.~\cite{ref:PF94}.
However, its influence on physical quantities other than energies
has not been explored sufficiently.

\textit{Framework.}
We implement the spherical Hartree-Fock-Bogolyubov (HFB) calculations
for the Pb nuclei.
The computational method is identical to that employed
in Ref.~\cite{ref:Nak13}.
We employ the M3Y-P6 semi-realistic interaction~\cite{ref:Nak13}
except the LS channels.
The Michigan-three-range-Yukawa (M3Y)-type interactions,
which were obtained from the $G$-matrix~\cite{ref:M3Y-P}
with phenomenological modification~\cite{ref:Nak13,ref:Nak03},
have been applied
in the mean-field (MF)~\cite{ref:Nak13,ref:Nak10b,ref:NS14}
and random-phase (RPA)~\cite{ref:Shi08} approximations.
A part of effects of the $3N$ interaction is included
by the density-dependence in the central channels,
which plays a crucial role in reproducing the saturation properties.
With the LS channels enhanced phenomenologically,
the M3Y-P6 parameter-set gives reasonable prediction of magic numbers
in wide range of the nuclear chart
including unstable nuclei~\cite{ref:NS14}.

The M3Y-type interaction has the LS channels of the following form:
\begin{equation}
v_{ij}^{(\mathrm{LS})} = \sum_n \big(t_n^{(\mathrm{LSE})} P_\mathrm{TE}
 + t_n^{(\mathrm{LSO})} P_\mathrm{TO}\big) f_n^{(\mathrm{LS})} (r_{ij})\,
 \mathbf{L}_{ij}\cdot(\mathbf{s}_i+\mathbf{s}_j)\,, \label{eq:LS}
\end{equation}
where the subscripts $i$ and $j$ are indices of nucleons,
$\mathbf{L}_{ij}= \mathbf{r}_{ij}\times \mathbf{p}_{ij}$
with $\mathbf{r}_{ij}= \mathbf{r}_i - \mathbf{r}_j$ and
$\mathbf{p}_{ij}= (\mathbf{p}_i - \mathbf{p}_j)/2$,
$\mathbf{s}_i$ is the spin operator, $r_{ij}=|\mathbf{r}_{ij}|$,
and $f_n^{(\mathrm{LS})}(r)=e^{-\mu_n^{(\mathrm{LS})} r}/\mu_n^{(\mathrm{LS})} r$.
$P_\mathrm{TE}$ ($P_\mathrm{TO}$) denotes the projection operator
on the triplet-even (triplet-odd) two-particle states.
In M3Y-P6, $v^{(\mathrm{LS})}$ of the original M3Y interaction derived
from the Paris $2N$ force~\cite{ref:M3Y-P}
was multiplied by an overall factor $2.2$,
to reproduce the s.p. level sequence in $^{208}$Pb.
On the other hand, Kohno has pointed out~\cite{ref:Koh12,ref:Koh13} that
the LS interaction effectively becomes stronger
as the nucleon density increases, because of the $3N$ interaction.
Following his result, we consider an alternative LS interaction
in which a density-dependent term is added,
\begin{eqnarray}
v_{ij}^{(\mathrm{LS}\rho)} &=& 2i\,D[\rho(\mathbf{R}_{ij})]\,
 \mathbf{p}_{ij}\times\delta(\mathbf{r}_{ij})\,\mathbf{p}_{ij}\cdot
 (\mathbf{s}_i+\mathbf{s}_j) \nonumber\\
&=& D[\rho(\mathbf{R}_{ij})]\,
 \Big(-\nabla_{ij}^2\delta(\mathbf{r}_{ij})\Big)\,
 \mathbf{L}_{ij}\cdot(\mathbf{s}_i+\mathbf{s}_j)\,, \label{eq:DDLS}
\end{eqnarray}
instead of enhancing $v^{(\mathrm{LS})}$.
Here $\rho(\mathbf{r})$ stands for the isoscalar nucleon density
and $\mathbf{R}_{ij}=(\mathbf{r}_i+\mathbf{r}_j)/2$.
The zero-range form of the $2N$ LS interaction is recovered
if $D[\rho]$ is replaced by a constant~\cite{ref:VB72}.
Conversely, density-dependence in $D[\rho]$ may represent
effects of the $3N$ interaction.
For functional form of $D[\rho]$,
the chiral $3N$ result suggests linear increase
at low density~\cite{ref:Koh-pv}.
We here take
\begin{equation}
D[\rho(\mathbf{r})] = -w_1\,\frac{\rho(\mathbf{r})}
 {1+d_1\rho(\mathbf{r})}\,, \label{eq:DinLS}
\end{equation}
with $w_1,d_1>0$.
This $D[\rho]$ is similar to the form found in Ref.~\cite{ref:BVB13}
for the central channel.
The second term of the denominator,
which guarantees $|D|<w_1/d_1$ even at extremely high density,
is used to avoid instability.
We adopt $d_1=1.0\,\mathrm{fm}^3$,
with which the $d_1\rho$ term remains small at $\rho\lesssim\rho_0$
($\rho_0$ denotes the saturation density).
We have confirmed that the results are insensitive to $d_1$.

For spherical nuclei,
contribution of $v^{(\mathrm{LS}\rho)}$ to the Hartree-Fock energy is given by
\begin{eqnarray}
E_\mathrm{HF}^{(\mathrm{LS}\rho)} &=& \frac{1}{4}\int d^3r\,D[\rho(\mathbf{r})]\,
 \nonumber\\
 &&\times \Big\{\rho(\mathbf{r})\,\nabla\cdot\mathbf{J}(\mathbf{r})
 + \sum_{\tau=p,n} \rho_\tau(\mathbf{r})\,\nabla\cdot\mathbf{J}_\tau(\mathbf{r})
 \nonumber\\
&&\quad - \mathbf{J}(\mathbf{r})\cdot\nabla\rho(\mathbf{r})
 - \sum_{\tau=p,n} \mathbf{J}_\tau(\mathbf{r})\cdot\nabla\rho_\tau(\mathbf{r})
 \Big\} \,,
 \label{eq:E_DDLS}
\end{eqnarray}
where
$\mathbf{J}_\tau(\mathbf{r})=2i\sum_{i\in\tau}\varphi_i^\dagger(\mathbf{r})\,
 \mathbf{s}_i\times\nabla\varphi_i(\mathbf{r})
=(\mathbf{r}/r^2)
\sum_{i\in\tau}\varphi_i^\dagger(\mathbf{r})\,
(2\mbox{\boldmath$\ell$}\cdot\mathbf{s}_i)\,\varphi_i(\mathbf{r})$
and $\mathbf{J}(\mathbf{r})=\sum_{\tau=p,n} \mathbf{J}_\tau(\mathbf{r})$,
with the s.p. function $\varphi_i(\mathbf{r})$
and $\mbox{\boldmath$\ell$}=\mathbf{r}\times\mathbf{p}$.
$E_\mathrm{HF}^{(\mathrm{LS}\rho)}$ yields
$\ell s$ terms of the s.p. potential,
\begin{eqnarray}
&&-\frac{1}{2r}\bigg[\,D[\rho(r)]\,
 \frac{d}{dr}\Big(\rho(r)+\rho_\tau(r)\Big) \nonumber\\
&&\quad + \frac{1}{2}\,\frac{\delta D}{\delta\rho}[\rho(r)]\,
 \Big(\rho(r)+\rho_\tau(r)\Big)\,\frac{d\rho(r)}{dr}\bigg]\,
 \mbox{\boldmath$\ell$}\cdot\mathbf{s}\,.\quad(\tau=p,n) \nonumber\\
&&\label{eq:lspot}
\end{eqnarray}
The rearrangement term that contains $\delta D/\delta\rho=-w_1/(1+d_1\rho)^2$
enhances effects of the $3N$ interaction on the $\ell s$ splitting,
as argued in Ref.~\cite{ref:Koh13}.

In the MF calculations so far,
the $2N$ LS interaction has been determined
by fitting the $\ell s$ splitting to the experimental data.
To investigate effects of the density-dependent LS channel
without influencing the $\ell s$ splitting,
we determine the parameter $w_1$ so as to reproduce the M3Y-P6 result
of the $n0i_{13/2}$-$n0i_{11/2}$ splitting at $^{208}$Pb.
This LS-modified variant of M3Y-P6 will be called M3Y-P6a.
As well as the $n0i$ splitting,
the s.p. energies do not change from those of M3Y-P6 significantly.
Influence on the binding energy is insignificant as well;
$0.4\,\mathrm{MeV}$ increase at $^{16}$O
and decrease by $4.9\,\mathrm{MeV}$ at $^{208}$Pb.
The parameters in $v^{(\mathrm{LS})}$ and $v^{(\mathrm{LS}\rho)}$
are tabulated in Table~\ref{tab:param-LS}.

\begin{table}
~\vspace*{-2.3cm}
\begin{center}
\caption{Parameters of the LS channels in M3Y-P6 and M3Y-P6a.
\label{tab:param-LS}}
\begin{tabular}{ccr@{.}lr@{.}lr@{.}l}
\hline\hline
parameters && \multicolumn{2}{l}{~~M3Y-P6} &
 \multicolumn{2}{l}{~~M3Y-P6a} \\
 \hline
$1/\mu_1^{(\mathrm{LS})}$ &(fm)& $0$&$25$ & $0$&$25$ \\
$t_1^{(\mathrm{LSE})}$ &(MeV)&~$-11222$&$2$ &~~$-5101$& \\
$t_1^{(\mathrm{LSO})}$ &(MeV)& $-4173$&$4$ & $-1897$& \\
$1/\mu_2^{(\mathrm{LS})}$ &(fm)& $0$&$40$ & $0$&$40$ \\
$t_2^{(\mathrm{LSE})}$ &(MeV)& $-741$&$4$ & $-337$& \\
$t_2^{(\mathrm{LSO})}$ &(MeV)& $-1390$&$4$ & $-632$& \\
$w_1$ &(MeV$\cdot$fm$^8$)& $0$& & $742$& \\
$d_1$ &(fm$^3$)& \multicolumn{2}{r}{---~~~~} & $1$& \\
\hline\hline
\end{tabular}
\end{center}
\end{table}

\textit{Results and discussions.}
We define the isotope shifts of the Pb nuclei
by $\mathit{\Delta}\langle r^2\rangle_p(\mbox{$^A$Pb})
=\langle r^2\rangle_p(\mbox{$^A$Pb})-\langle r^2\rangle_p(\mbox{$^{208}$Pb})$,
where $\langle r^2\rangle_p(\mbox{$^A$Pb})$ represents
the proton mean-square (m.s.) radius in the $^A$Pb nucleus ($A=N+82$).
$\mathit{\Delta}\langle r^2\rangle_p$ has been measured precisely
for a chain of the Pb nuclei~\cite{ref:AHS87,ref:Ang04},
disclosing a remarkable kink at $N=126$.

We here note the finite-size effect of the constituent protons
on the isotope shifts.
In the convolution model which is customarily used
for the charge density distribution,
contribution of the charge radius of constituent protons
is canceled out in $\mathit{\Delta}\langle r^2\rangle_p$
because the proton number $Z$ is fixed.

The previous MF calculations indicate
that the kink in $\mathit{\Delta}\langle r^2\rangle_p(\mbox{$^A$Pb})$ at $N=126$
occurs owing to occupation of the $n0i_{11/2}$ orbit~\cite{ref:RF95}.
As the s.p. function of $n0i_{11/2}$ has a larger radius
than those of the neighboring orbits,
the attraction between protons and neutrons makes $\langle r^2\rangle_p$ larger
as $n0i_{11/2}$ is occupied to greater degree.
Because the neutron occupation on $0i_{11/2}$ is negligible in $N\leq 126$
while sizable in $N>126$,
$\mathit{\Delta}\langle r^2\rangle_p$ has a kink at the magic number $N=126$.
Although neutrons primarily occupy $1g_{9/2}$ beyond $N=126$,
the pair correlation allows them to occupy $0i_{11/2}$.
The s.p. energy difference $\varepsilon_n(0i_{11/2})-\varepsilon_n(1g_{9/2})$
was considered important in the previous studies,
to which the occupation probability on $n0i_{11/2}$ is sensitive.
On this basis, it was argued
that the kink in $\mathit{\Delta}\langle r^2\rangle_p(\mbox{$^A$Pb})$
may be relevant to the isospin-dependence
of the $\ell s$ potential~\cite{ref:SLKR95}.
However, it has been difficult to give a kink comparable to the observed one
unless $n1g_{9/2}$ and $n0i_{11/2}$ are nearly degenerate~\cite{ref:RF95}.

\begin{figure}
\includegraphics[scale=0.6]{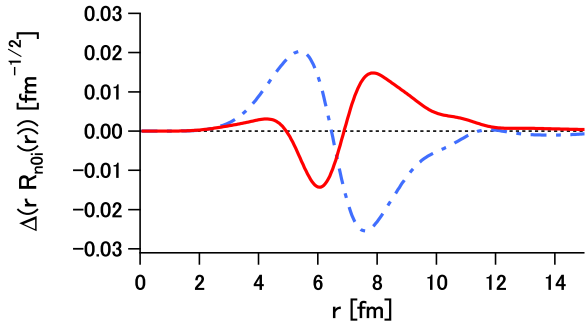}
\caption{(Color online) Difference of the radial part of the s.p. functions
for $n0i_{13/2}$ (blue dot-dashed line) and $n0i_{11/2}$ (red solid line);
$r\,R_{n\ell j}(r)$ obtained with M3Y-P6a relative to that with M3Y-P6,
by the HF calculations at $^{208}$Pb.
\label{fig:dspwf_n0i}}
\end{figure}

The density-dependence in the LS channel leads to an additional effect.
The LS channels of the $2N$ interaction
may derive the $\ell s$ potential whose strength is proportional
to the derivative of the density,
indicating that the $\ell s$ splitting is a surface effect.
Equation~(\ref{eq:DinLS}) gives the larger $|D[\rho]|$ for the higher $\rho$.
Therefore, when the size of the $\ell s$ splitting is equated,
$D[\rho]$ in Eq.~(\ref{eq:DDLS}) makes the $\ell s$ potential
stronger in the interior and weaker in the exterior,
as recognized from Eq.~(\ref{eq:lspot}).
Since the $\ell s$ potential acts attractively on the $j=\ell+1/2$ orbits
and repulsively on the $j=\ell-1/2$ orbits,
variational calculations shrink the wave functions of the $j=\ell+1/2$ orbits
while extend those of the $j=\ell-1/2$ orbits.
This trend is confirmed by difference of the radial functions $R_j(r)$
for $j=n0i_{13/2}$ and $n0i_{11/2}$ shown in Fig.~\ref{fig:dspwf_n0i},
which are obtained by the HF calculations at $^{208}$Pb.
We take the phase $R_{n0i}(r)\geq 0$ as usual.
The m.s. radius of the s.p. function of $n0i_{11/2}$
increases by $0.49\,\mathrm{fm}^2$
as we switch the interaction from M3Y-P6 to M3Y-P6a.
Note that the m.s. radius of $n1g_{9/2}$ does not differ much
between M3Y-P6 and M3Y-P6a.
The root-m.s. matter radius of the $^{208}$Pb nucleus
changes by $-0.03\,\mathrm{fm}$,
and the neutron-skin thickness
$\sqrt{\langle r^2\rangle_n}-\sqrt{\langle r^2\rangle_p}$
by $+0.001\,\mathrm{fm}$,
both of which are basically irrelevant to the isotope shifts.

\begin{figure}
\includegraphics[scale=0.45]{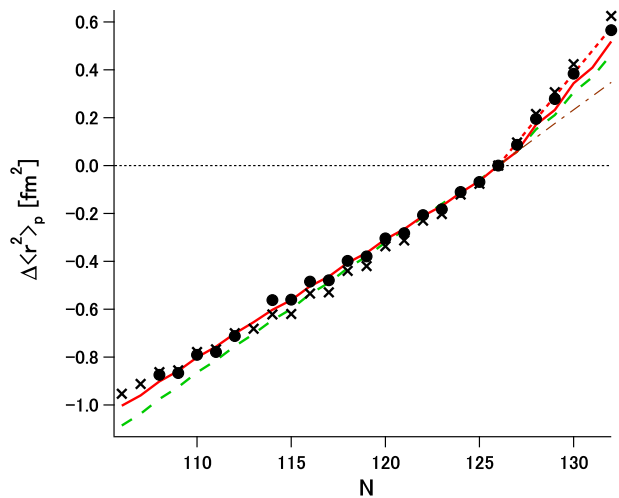}
\caption{(Color online) Isotope shifts of the Pb nuclei
$\mathit{\Delta}\langle r^2\rangle_p(\mbox{$^A$Pb})$,
obtained from the HFB calculations with M3Y-P6a (red solid line),
in comparison to those with M3Y-P6 (green dashed line).
Thin brown dot-dashed line in $N\geq 126$ is the HF result
in which all the valence neutrons occupy $n1g_{9/2}$.
Red dotted line shows $\mathit{\Delta}\langle r^2\rangle_p(\mbox{$^A$Pb})$
with M3Y-P6a in the hypothetical limit
that $n1g_{9/2}$ and $n0i_{11/2}$ were equally occupied.
Experimental data are taken from Refs.~\cite{ref:AHS87} (circles)
and \cite{ref:Ang04} (crosses).
\label{fig:Pb_drp}}
\end{figure}

We depict $\mathit{\Delta}\langle r^2\rangle_p$ in Fig.~\ref{fig:Pb_drp}.
The HFB results with M3Y-P6a (red solid line) are compared
to those with M3Y-P6 (green dashed line).
The kink at $N=126$ takes place because of the $n0i_{11/2}$ occupation;
indeed it becomes invisible
if all the valence neutrons in $N>126$ occupy $n1g_{9/2}$,
as shown by the thin brown dot-dashed line.
The larger $n0i_{11/2}$ radius provides the more rapid increase
of $\mathit{\Delta}\langle r^2\rangle_p$ in $N>126$.
Thus the stronger kink is obtained with M3Y-P6a than with M3Y-P6,
in better agreement with the experimental data~\cite{ref:AHS87,ref:Ang04}.
The same trend is obtained with M3Y-P7~\cite{ref:Nak13} and D1M~\cite{ref:D1M}
after replacing a part of the LS channel by $v^{(\mathrm{LS}\rho)}$
of Eq.~(\ref{eq:DDLS}).

\begin{figure}
\includegraphics[scale=0.4]{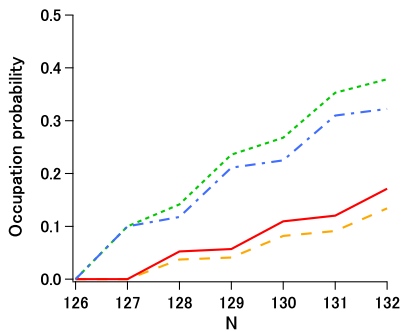}
\caption{(Color online) Occupation probabilities
on $n1g_{9/2}$ and $n0i_{11/2}$ obtained by the HFB calculations.
Blue dot-dashed (green dotted) line is for $n1g_{9/2}$
and red solid (orange dashed) line for $n0i_{11/2}$
in the M3Y-P6a (M3Y-P6) results.
\label{fig:Pb_occ-prob}}
\end{figure}

In the limit that $n0i_{11/2}$ has equal occupation probability
to the lower-lying $n1g_{9/2}$ orbit,
the s.p. functions obtained by M3Y-P6
give $\mathit{\Delta}\langle r^2\rangle_p$
indistinguishable from the red solid line in Fig.~\ref{fig:Pb_drp},
and therefore comparable to the observed one in $N>126$.
This is analogous to the results reported in the previous studies
without density-dependence in the LS channels~\cite{ref:RF95,ref:SLR94}.
However, the energy difference between $n1g_{9/2}$ and $n0i_{11/2}$
is not negligible in the experimental data,
being $0.78\,\mathrm{MeV}$
if extracted from the lowest states of $^{209}$Pb~\cite{ref:TI}.
With this difference it is unlikely that the occupation probabilities
of these orbits are so close.
In contrast,
with M3Y-P6a $n0i_{11/2}$ lies higher than $n1g_{9/2}$ by $0.72\,\mathrm{MeV}$
at $^{208}$Pb,
and the occupation probability on $n0i_{11/2}$ is substantially lower
than that on $n1g_{9/2}$,
as presented in Fig.~\ref{fig:Pb_occ-prob}.
Nevertheless, owing to the density-dependence in the LS channel,
the kink of $\mathit{\Delta}\langle r^2\rangle_p$
is reproduced to fair degree.
If these orbits were equally occupied in $N>126$,
$\mathit{\Delta}\langle r^2\rangle_p$ could increase further,
providing the red dotted line in Fig.~\ref{fig:Pb_drp}.

The even-odd staggering of $\mathit{\Delta}\langle r^2\rangle_p(\mbox{$^A$Pb})$
observed in $N\geq 126$ is well reproduced by the HFB calculations.
This originates from the staggering in the occupation probabilities
in Fig.~\ref{fig:Pb_occ-prob},
which owes to a quasiparticle on $n1g_{9/2}$ in the ground states
of the odd-$N$ isotopes.
Thus the even-odd staggering supports the picture
that $n0i_{11/2}$ is responsible
for the kink of $\mathit{\Delta}\langle r^2\rangle_p$.

In Fig.~\ref{fig:Pb_drp}, we find that
increase of $\mathit{\Delta}\langle r^2\rangle_p(\mbox{$^A$Pb})$ in $N\leq 126$
is slower with M3Y-P6a than with M3Y-P6,
which is linked to the smaller radius of $n0i_{13/2}$ with M3Y-P6a.
The slope of $\mathit{\Delta}\langle r^2\rangle_p$ in the M3Y-P6a results
is in good agreement with the data in wide region of $N\leq 126$,
although the even-odd staggering is not visible as in the data.

We thus confirm that the isotope shifts of the Pb nuclei
can be described by the density-dependent LS interaction
without the $n1g_{9/2}$-$n0i_{11/2}$ degeneracy.
As such density-dependence is naturally derived from the $3N$ interaction
as shown by Kohno,
the strong kink in the isotope shifts may be regarded
as an effect of the $3N$ interaction.
Importance of the $3N$ interaction in the $\ell s$ splitting
was pointed out also in Ref.~\cite{ref:LS-MC}.
On the contrary, it was suggested~\cite{ref:LS-UMOA}
that many-body correlations induced by the $2N$ interaction
could enhance the $\ell s$ splitting.
Although the many-body correlations do not yield
significant density-dependence
within the lowest-order Brueckner theory~\cite{ref:Koh12},
their effects should be investigated further
for complete understanding.
We remark that the effects of the $3N$ interaction presented here
are on the wave functions, not on the energies,
and are therefore unrenormalizable
with usual density-independent $2N$ interactions.

We finally comment on the strength of the LS interaction.
As is important to nuclear structure,
the $\ell s$ splitting should not change much
even when density-dependence is introduced.
All the effects of the density-dependent LS interaction presented in this paper
become the stronger with the greater $w_1$ in Eq.~(\ref{eq:DinLS}),
when keeping the $\ell s$ splitting
by enhancing or reducing the $2N$ LS strength.
In the present study, $w_1$ has been fixed
so as not to alter the $\ell s$ splitting of M3Y-P6,
which gives reasonable shell structure,
while the $2N$ LS interaction is returned to that of the M3Y-Paris interaction.
In Ref.~\cite{ref:Koh12}, Kohno showed
that the effective strength of the LS interaction at $\rho\approx\rho_0$
deduced from the chiral $3N$ interaction~\cite{ref:EGM05}
is comparable to the empirical strength
that reproduces the observed $\ell s$ splitting,
though significantly weak at lower densities.
However, we have found that the $\ell s$ splitting is better connected
to the strength at $\rho\approx(2/3)\rho_0$, rather than at $\rho=\rho_0$.
It should also be noted that the current $\chi$EFT~\cite{ref:EGM05}
(so-called N$^3$LO) is not guaranteed
to be fully convergent at $\rho\sim\rho_0$.
We have therefore used the functional form of Eq.~(\ref{eq:DDLS})
which is consistent
with the $\chi$EFT calculations~\cite{ref:Koh12,ref:Koh13,ref:Koh-pv},
but have not taken the strength derived in Ref.~\cite{ref:Koh12,ref:Koh13}
seriously,
emphasizing the qualitative effects of the $3N$ interaction.

\textit{Summary.}
We have investigated effects of the density-dependent LS interaction,
which has been derived from the chiral $3N$ interaction by Kohno,
on the isotope shifts of the Pb nuclei.
As this LS interaction effectively comes stronger as the density grows,
the wave functions of the $j=\ell+1/2$ orbits shrink
while those of the $j=\ell-1/2$ orbits distribute more broadly.
Since the attraction from neutrons occupying $0i_{11/2}$
produces the kink in the isotope shift of the Pb nuclei at $N=126$,
the broader $n0i_{11/2}$ function makes the kink stronger.
By introducing density-dependence in the LS channel
with the strength so as to keep the original $\ell s$ splitting,
we can reproduce the data on the isotope shifts of the Pb nuclei
with the Hartree-Fock-Bogolyubov calculations
in a long chain of neutron numbers,
as has been exemplified by an LS-modified variant
of the semi-realistic M3Y-P6 interaction.
It is remarked that the rapid increase of the radius in $N>126$
can be described fairly well
without degeneracy between the $n1g_{9/2}$ and $n0i_{11/2}$ levels,
unlike previous studies.

Further exploration of effects of the density-dependent LS interaction
will be desired,
including application to the deformed MF calculations
as well as to the RPA calculations.

\begin{acknowledgments}
%
Discussion with M.~Kohno is gratefully acknowledged.
This work is financially supported in part
as KAKENHI No.~24105008 by The MEXT, Japan,
and as No.~25400245 by JSPS.
Numerical calculations are performed on HITAC SR16000s
at IMIT in Chiba University,
ITC in University of Tokyo,
IIC in Hokkaido University,
and YITP in Kyoto University.
\end{acknowledgments}


\end{document}